\documentclass[a4paper]{amsproc}

\usepackage{pstricks}
\usepackage{pstcol}
\usepackage{graphicx}
\usepackage{amsmath}
\usepackage{amsfonts}
\usepackage{amssymb}
\usepackage{latexsym}

\newtheorem{lemma}{Lemma}
\newtheorem{theorem}{Theorem}

\newcommand{\R}{\text{\rm Re }}
\newcommand{\I}{\text{\rm Im }}
\newcommand{\nb}{\text{\bf n}}

\newcommand{\Diff}{\text{\rm Diff }}

\newcommand{\Vect}{\text{\rm Vect }}

\begin{document}
\title[Energy characteristics of subordination chains]
{Energy characteristics of subordination chains}
\author[Alexander Vasil'ev]{Alexander Vasil'ev}
\address{Matematisk institutt, Universitetet i Bergen, Johannes Brunsgate 12, N-5008, Bergen, Norway} \email{alexander.vasiliev@uib.no} 
\begin{abstract}
We consider subordination chains of simply connected domains with smooth boundaries 
in the complex plane.
Such chains admit Hamiltonian and Lagrangian interpretations through the L\"owner-Kufarev
evolution equations. The action functional is constructed and its time variation is obtained.
It represents the infinitesimal 
version of the action of the Virasoro-Bott group over the space of analytic univalent
functions.
\end{abstract}

\maketitle

\section{Introduction}

Many physical processes may be interpreted as expanding dynamical systems of
domains in the complex plane $\mathbb C$ or in the Riemann sphere $\hat{\mathbb C}$. This leads
to the study of time-parameter L\"owner subordination chains. In particular, we are interested
in L\"owner chains of simply connected univalent domains with smooth ($C^{\infty}$) boundaries.
There exists a canonical identification of the space of such domains (under certain conformal normalization) with the infinite dimensional K\"ahler manifold whose central extension is the Virasoro-Bott group.
A L\"owner subordination chain $\Omega(t)$ is described by time-dependent family of conformal maps $z=f(\zeta,t)$
from the unit disk $U$ onto $\Omega(t)$, normalized by $f(\zeta,t)=a_1(t)\zeta+a_2(t)\zeta^2+\dots$,
$a_1(t)>0$. After 1923 seminal L\"owner's paper \cite{Loewner} a fundamental contribution to
the theory of L\"owner chains has been made by Pommerenke \cite{Pommerenke1, Pommerenke2} who
described governing evoluton equations in partial and ordinary derivatives, known now as
the L\"owner-Kufarev equations due to Kufarev's work \cite{Kufarev}. A particular case
of subordination dynamics is presented by the Laplace growth evolution (or Hele-Shaw advancing evolution) 
consisting of the Dirichlet problem for a harmonic potential where the boundary of the phase domain is unknown {\it a priori} (free boundary), and, in fact, is defined by the normality of its motion.
(see, e.g., \cite{Howison, Richardson, Vasiliev1}). The aim of our paper is to give a Hamiltonian and Lagrangian descriptions of the subordination evolution. In particular, we discuss the relations
between the L\"owner-Kufarev equations in partial and ordinary derivatives, construct the action functional,
obtain its time variation. This variation  represents the infinitesimal 
version of the action of the Virasoro-Bott group over the space of analytic univalent
functions.

\section{Hamiltonian formulation of the subordination evolution}

The parametric method  emerged more than 80 years ago in the
celebrated paper by L\"owner \cite{Loewner} who studied a
time-parameter semigroup of conformal one-slit maps of the unit disk $U$ coming
then at an evolution equation called after him. His main
achievement was an infinitesimal description of a semi-flow of
such maps by the Schwarz kernel that led him to the L\"owner
equation. This crucial result was generalized, then, in several
ways (see \cite{Pommerenke2} and the references therein).

A time-parameter family $\Omega(t)$ of simply connected hyperbolic univalent domains forms a {\it subordination chain}  in the complex plane $\mathbb
C$, for $0\leq t< \tau$ (where $\tau$ may be $\infty$), if
$\Omega(t)\subset \Omega(s)$,  $\Omega(t)\neq \Omega(s)$, whenever $t<s$.
We
suppose that the origin is an interior point of the kernel of
$\{\Omega(t)\}_{t=0}^{\tau}$, and the boundaries $\partial \Omega(t)$ are smooth ($C^{\infty}$). Let us normalize the growth of the
evolution of this subordination chain by the conformal radius of
$\Omega(t)$ with respect to the origin to be $e^t$. By the Riemann
Mapping Theorem we construct a subordination chain of mappings
$f(\zeta,t)$, $\zeta\in U$,  where each function $\displaystyle
f(\zeta,t)=e^t\zeta+a_2(t)\zeta^2+\dots$ is a holomorphic univalent
map of $U$ onto $\Omega(t)$ for every fixed $t$.

Pommerenke's
result \cite{Pommerenke1, Pommerenke2} says that given a subordination
chain of domains $\Omega(t)$ defined for $t\in [0,\tau)$, there exists
an analytic regular function
$$p(\zeta,t)=1+p_1(t)\zeta+p_2(t)\zeta^2+\dots,\quad
\zeta\in U,$$ such that $\R p(\zeta, t)>0$  and
\begin{equation}
\frac{\partial f(\zeta,t)}{\partial t}=\zeta\frac{\partial
f(\zeta,t)}{\partial \zeta}p(\zeta,t),\label{LK}
\end{equation}
for $\zeta\in U$ and for almost all $t\in [0,\tau)$.  The
equation (\ref{LK}) is called the L\"owner-Kufarev equation due to
two seminal papers by L\"owner \cite{Loewner} with
\begin{equation}
p(\zeta,t)=\frac{e^{iu(t)}+\zeta}{e^{iu(t)}-\zeta},\label{yadro}
\end{equation}
where $u(t)$ is a continuous function regarding to $t\in [0,\tau)$,
 and by Kufarev \cite{Kufarev} in general case,
where this equation appeared for the first time.

In \cite{Vasiliev2}, the case of smooth boundaries $\partial \Omega(t)$,
being embedded into the class of quasidisks, has been proved to admit
a specific integral form of the function $p(\zeta,t)$ as
\[
p(\zeta,t)=1+\frac{\zeta}{2\pi i}\int\limits_{S^1=\partial U}\frac{\nu(\omega,t)}{\omega(\omega-\zeta)}d\omega,
\]
for almost all $t\in [0,\tau]$, where the function $\nu(\omega,t)$ belongs to the Lie algebra $\Vect S^1$ of the vector fields on the unit circle $S^1$, with the the Poisson - Lie
bracket given by
\[[\nu_1,\nu_2]={\nu}_1{\nu}'_2-{\nu}_2{\nu}'_1,\]
where the derivatives are taken with respect to the angle variable of $S^1$. Comparing
with the Herglotz representation
\[
p(\zeta,t)=\zeta\int\limits_{S^1}\frac{\omega+\zeta}{\omega-\zeta}d\mu(\omega,t),
\]
for the family of Herglotz measures normalized as $\int_{S^1}d\mu(\omega,t)=1$, we deduce that $d\mu(\omega,t)=\rho(\omega,t)|d\omega|$,
$\omega\in S^1$, and
\[\rho(e^{i\theta},t)\equiv \frac{1}{4\pi}\nu(e^{i\theta},t).\]
From the other side, $\R p(e^{i\theta},t)=2\pi\rho(e^{i\theta},t)$ and the real-valued function
$\rho(e^{i\theta},t)$ is non-negative for almost all $t\in [0,\tau)$.

To arrive at the Hamiltonian interpretation of subordination dynamics, let us 
rewrite equation (\ref{LK}) in the form
\begin{equation}
\frac{\partial f'(\zeta,t)}{\partial t}=\frac{\partial H(\zeta,f',t)}{\partial \zeta},\label{H1}
\end{equation}
where $H(\zeta,f',t)=\zeta f'(\zeta,t) p(\zeta,t)$, and the derivative $f'$ is taken with respect to the complex variable $\zeta$. Interpreting the function $H(\zeta,f',t)$ as a Hamiltonian we must write
\begin{equation}
\frac{\partial \zeta}{\partial t}=-\frac{\partial H(\zeta,f',t)}{\partial f'}=-\zeta p(\zeta,t),\label{H2}
\end{equation}
formally yet. Equations (\ref{H1}) and (\ref{H2}) constitute the conjugate pair of Hamilton's relations,
however this requires some additional clearance to give sense to the equation (\ref{H2}). This equation
is just the L\"owner-Kufarev equation in ordinary derivatives $\dot{\zeta}=-\zeta p(\zeta,t)$.

The equation (\ref{LK}) represents a growing evolution of simply
connected domains. Let us consider the reverse process. Given an
initial domain $\Omega(0)\equiv \Omega_0$ (and therefore, the
initial mapping $f(\zeta,0)\equiv f_0(\zeta)$), and a function
$p(\zeta,t)$ with positive real part normalized by $p(\zeta,t)=1+p_1\zeta+\dots$, we solve the equation (\ref{LK})
and ask whether the solution $f(\zeta,t)$ represents a subordination
chain of simply connected domains. The initial condition
$f(\zeta,0)=f_0(\zeta)$ is not given on the characteristics of the
partial differential equation (\ref{LK}), hence the solution exists
and is unique. Assuming $s$ as a parameter along the characteristics
we have $$ \frac{dt}{ds}=1,\quad \frac{d\zeta}{ds}=-\zeta
p(\zeta,t), \quad \frac{df}{ds}=0,$$ with the initial conditions
$t(0)=0$, $\zeta(0)=z$, $f(\zeta,0)=f_0(\zeta)$, where $z$ is in
$U$.  Obviously, $t=s$. We still need to give sense to this formalism
because the domain of $\zeta$ is the entire unit disk, however the solutions to
the second equation of the characteristic system range within the unit disk but do not fill it. 
Therefore, introducing another letter $w$ in order to distinct the function $w(z,t)$  from the variable $\zeta$, we arrive at the Cauchy problem for the  L\"owner-Kufarev
equation in ordinary derivatives for a function $\zeta=w(z,t)$
\begin{equation}
\frac{dw}{dt}=-wp(w,t),\label{LKord}
\end{equation}
with the initial condition $w(z,0)=z$. The equation (\ref{LKord}) is the non-trivial  characteristic
equation for (\ref{LK}). Unfortunately, this approach requires the
extension of $f_0(w^{-1}(\zeta,t))$ into $U$ ($w^{-1}$ means the inverse function) because the solution to
(\ref{LK}) is the function $f(\zeta,t)$  given as
$f_0(w^{-1}(\zeta,t))$, where $\zeta=w(z,s)$ is a solution of the
initial value problem for the characteristic equation (\ref{LKord}) (or (\ref{H2}))
that maps $U$ into $U$. Therefore, the solution of the initial
value problem for the equation (\ref{LK}) may be non-univalent.

Let $A$ stand for the usual class of all univalent holomorphic
functions $f(z)=z+a_2z^2+\dots$ in the unit disk. Solutions to the
equation (\ref{LKord}) are regular univalent functions
$w(z,t)=e^{-t}z+a_2(t)z^2+\dots$ in the unit disk that map $U$  into
itself. Conversely, every function from the class $A$ can be
represented by the limit
\begin{equation}
f(z)=\lim\limits_{t\to\infty}e^t w(z,t),\label{limit}
\end{equation}
where there exists a function $p(z,t)$ with positive real part for almost
all $t\geq 0$, such that $w(z,t)$ is a solution to the eqiation
(\ref{H2}) (see \cite[pages 159--163]{Pommerenke2}). Each function
$p(z,t)$ generates a unique function from the class $A$. The
reciprocal statement is not true. In general, a function $f\in A$
can be determined by different functions $p$.

From \cite[page 163]{Pommerenke2} it follows that we can guarantee the
univalence of the solutions to the L\"owner-Kufarev equation in
partial derivatives (\ref{LK}) assuming the initial condition
$f_0(\zeta)$ given by the limit (\ref{limit}) with the function
$p(\cdot,t)$ chosen to be the same in the equations (\ref{LK}) and
(\ref{LKord}). Originally, these arguments have been made by Prokhorov and the author in
\cite{ProkhVas}. We remark also that an analogous Hamiltonian $H$ one may also find in \cite{Blumenfeld}.

The Hamiltonian $H$ is linear with respect to the variable $f'$, therefore, the Hamiltonian
dynamics which is generated by $H$ is trivial and the velocity is constant. In \cite{ProkhVas}
we studied another Hamiltonian system for a finite number of the coefficients of the function $w(z,t)$ generated by the equation (\ref{LKord}). It turns out that the Hamiltonian system generated
by the coefficients is Liouville partially integrable and the first
integrals were obtained and were proved to possess a contact structure. However,
the Hamiltonian was again linear with respect to the conjugate system, and we have both
systems accelerationless. In order to describe a non-trivial motion we proceed with the Lagrangian
formulation.

\section{Lagrangian formulation of the subordination evolution}

Let us consider a subordination chain $\{\Omega(t)\}_{t=0}^{\tau}$, $0\in \Omega(t)$, and the
time-parameter family of the real-valued Green functions $G(z,t)$ of $\Omega(t)$ with the logarithmic singularity at $0$.
If $z=f(\zeta,t)$ is the Riemann map from the unit disk $U$ onto $\Omega(t)$, $f'(0,t)=e^t$, then
$G(z,t)=-\log |f^{-1}(z,t)|$. The unit normal vector $\nb$ to $\partial \Omega(t)$ in the outward direction
can be written as
\[ 
\nb=\frac{\zeta f'(\zeta,t)}{|f'(\zeta,t)|},\quad |\zeta|=1.
\]
Therefore, the normal velocity $v_n$ of the boundary $\partial \Omega(t)$ at the point $f(e^{i\theta},t)$ may be expressed as
\[
v_n=\R \dot{f}\,\,\overline{\frac{e^{i\theta} f'}{|f'|}}=|f'|\R \frac{\dot{f}}{e^{i\theta} f'}=
|f'|\rho(e^{i\theta},t),
\]
where the function $\rho(e^{i\theta},t)$ was defined in the preceding section. Thus, $\rho(e^{i\theta},t)=v_n |\nabla G|$.

The easiest Lagrangian is given by the Dirichlet integral
\[
\iint\limits_{D}|\nabla G|^2d\sigma_z,
\]
where $d\sigma_z= \big|\frac{dz\wedge \, d\bar{z}}{2} \big|$,  locally for any measurable set $D\subset \Omega(t)\setminus\{0\}$.
However, this functional cannot be defined globally in
$\Omega(t)\setminus\{0\}$ because of the parabolic singularity at
the origin. To overcome this obstacle we define the energy represented by this Lagrangian
in the following way. Let
$\Omega_{\varepsilon}(t)=\Omega(t)\setminus
\{z:\,|z|\leq\varepsilon\}$ for a sufficiently small $\varepsilon$,
$U_{\varepsilon}=\{\zeta:\,\varepsilon<|\zeta|<1\}$.
Then the finite limit
\begin{equation}
\mathcal{E}=\lim\limits_{\varepsilon\to
0}\left\{\iint\limits_{\Omega_{\varepsilon}(t)}|\nabla G|^2
d\sigma_z+2\pi\log\varepsilon\right\}\label{action1}
\end{equation}
exists. Applying the conformal map $z=f(\zeta,t)$ and changing variables we arrive at the representation
of the energy
\[\mathcal{E}=\mathcal{E}[f]= 2\pi\log |f'(0,t)|=2\pi t,\]
just by the capacity (or the conformal radius in this case) of $\partial\Omega(t)$. In other words,
$\mathcal{E}$ represents the classical action for the Lagrangian defined by the Dirichlet integral.
This interpretation allows us to get less trivial Lagrangian description of subordination dynamics,
that in particular, emerges in the Liouville part of the CFT \cite{Polyakov, Takhtajan}.

The classical field theory studies the extremum of the action
functional, and its critical value is called the classical
action. The critical point $\phi^*$ satisfies  Hamilton's principle
(or the principle of the least action), i.e., $\delta
\mathcal{S}[\phi^*]=0$, which is the Euler-Lagrange equation for the
variational problem defined by the action functional $\mathcal{S}[\phi]$. For the action
given by the Dirichlet integral the classical action is achieved for
the harmonic $\phi^*$ and the principle of the least action leads to
the Laplacian equation $\Delta \phi=0$ as above. The Liouville action plays a key role in two-dimensional
gravity and leads to the Liouville equation, a solution of which is the Poincar\'e
metric of the constant negative curvature. This conformal metric is of
a paticular interest, because no flat metric satisfies the Einstein Field Equation.
The conformal symmetry of CFT is generated by its energy-momentum tensor $T$ whose mode expansion
is expressed in terms of the operators satisfying the commutation relations of the Virasoro algebra.
The (2,0)-component of the energy-momentum tensor in the Liouville theory is given by the expression
$T_{\varphi}=\varphi_{zz}-\frac{1}{2}\varphi_z^2$ that leads to the classical Schwarz result $T_{\varphi}=S_f(\zeta)$
with the Schwarzian derivative $S_f$ where $f$ is the ratio of two linearly independent solutions
to the Fuchsian equation $w''+\frac{1}{2}T_{\varphi}w=0$.

In the case of subordination chains our starting point will be a Riemannian metric
$ds^2=e^{\varphi (z)}|dz|^2$. In the case of the Liouville theory, the real-valued potential $\varphi$
satisfies the Liouville equation $\varphi_{z\bar{z}}=\frac{1}{2}e^{\varphi}$ (generally, with
certain prescribed asymptotics which guarantee the uniqueness). Geometrically, this means
that the conformal metric $ds^2$ has constant negative curvature -1 on the underlying Riemann surface
corresponding to prescribed singularities.  Let us consider
the complex Green function $W(z,t)$ whose real part is $G(z,t)=\log |f^{-1}(z,t)|$, $z\in \Omega(t)\setminus 0$, as before. We have the
representation $W(z,t)=-\log z+w_0(z,t)$, 
where $w_0(z,t)$ is an analytic regular function in $\Omega(t)$.
Because of the conformal invariance of the Green function we have the
superposition
\[(W\circ f)(\zeta,t)=-\log\,\zeta,\]
and the conformally invariant complex velocity field is just
$W'(z,t)=-\frac{{f^{-1}}'}{f^{-1}}(z,t)$, where $\zeta=f^{-1}(z,t)$
is the inverse to our parametric function $f$ and prime means the
complex derivative. Rewriting this relation we get
\begin{equation}
(W'(z,t)\,dz)^2=\frac{d\zeta^2}{\zeta^2}.\label{qudratic}
\end{equation}
The velocity field is the conjugation of $(-W')$. In
other words the velocity field is directed along the trajectories of
the quadratic differential in the left-hand side of (\ref{qudratic})
for each fixed moment $t$. The equality (\ref{qudratic}) implies
that the boundary $\partial\Omega(t)$ is the orthogonal trajectory
of the differential $(W'(z,t)\,dz)^2$ with a double pole at the
origin. The dependence on $t$ yields that the trajectory structure
of this differential changes in time, and in general, the stream
lines are not inherited in time. These lines are geodesic in the
conformal metric $|W'(z,t)||dz|$ generated by this differential. 
Let us use the conformal logarithmic metric generated by
(\ref{qudratic}) 
\[
ds^2=\frac{|{f^{-1}}'|^2}{|f^{-1}|^2}|dz|^2=|W'|^2|dz|^2,
\]
which is intrinsically flat.
Unlike the Poincar\'e metric, the hyperbolic boundary
is not singular for the logarithmic metric whereas the origin is. But it is a
parabolic singularity which can be easily regularized. The density of this
metric satisfies the usual Laplacian equation $\varphi_{z\bar{z}}=0$
in $\Omega(t)\setminus\{0\}$, where
$\varphi(z)=\log\frac{|{f^{-1}}'|^2}{|f^{-1}|^2}$.
The function
$\varphi$ possesses the asymptotics
\[
\varphi\sim \log \frac{1}{|z|^2},\quad  |\varphi_z|\sim
\frac{1}{|z|^2} \quad \mbox{as $z\to 0$},
\]
therefore, the finite limit
\begin{equation}
\mathcal{S}=\mathcal{S}[\varphi]=\lim\limits_{\varepsilon\to
0}\left\{\int\limits_{\Omega_{\varepsilon}(t)}|\varphi_z|^2
d\sigma_z+2\pi\log\varepsilon\right\}\label{action2}
\end{equation}
exists and is called the logarithmic action.

\begin{lemma}
The Euler-Lagrange equation for the variational problem for the
logarithmic action $\mathcal{S}[\phi]$ is the Laplacian equation
$\Delta\phi=-4\pi\delta_0(z)$, $z\in \Omega(t)$, where $\delta_0(z)$
is the Dirac distribution supported at the origin, where $\phi$ is
taken from the class of twice differentiable functions in
$\Omega(t)\setminus \{0\}$ with the asymptotics $\phi\sim -\log
|z|^2$ as $z\to 0$.
\end{lemma}
\begin{proof}
Let us consider first the integral
\[
\mathcal{S}_{\varepsilon}[\phi]=\int\limits_{\Omega_{\varepsilon}(t)}|\phi_z|^2
d\sigma_z=\int\limits_{\mathbb
C}\chi_{\Omega_{\varepsilon}(t)}|\phi_z|^2 d\sigma_z,
\]
where $\chi_{\Omega_{\varepsilon}(t)}$ is the characteristic
function of $\Omega_{\varepsilon}(t)$. Then, due to Green's theorem,
\begin{eqnarray}
\lim\limits_{h\to
0}\frac{\mathcal{S}_{\varepsilon}[\phi+hu]-\mathcal{S}_{\varepsilon}[\phi]}{h}
& =& 2\int\limits_{\mathbb C}\chi_{\Omega_{\varepsilon}(t)}\R
\phi_z\overline{u_z}\,d\sigma_z\nonumber\\
&=& -\frac{1}{2}\int\limits_{\Omega_{\varepsilon}(t)}u\Delta\phi\,
d\sigma_{z}+ \frac{1}{2}\int\limits_{\partial
\Omega_{\varepsilon}(t)}u\frac{\partial \phi}{\partial
n}\,ds,\label{limit1}
\end{eqnarray}
in distributional sense for every $C^{\infty}(\mathbb C)$ test
function $u$ supported in $\Omega(t)$. On the other hand, we have
$\partial\phi/\partial n\sim -2/\varepsilon$ as $\varepsilon\to 0$
and $u=0$ on $\partial \Omega(t)$. Therefore, the expression
(\ref{limit1}) tends to
\[
-\frac{1}{2}\int\limits_{\Omega(t)}u\Delta\phi d\sigma_{z}-2\pi
u(0),
\]
as $\varepsilon\to 0$, and the latter must vanish, that is
equivalent to the Laplacian equation mentioned in the statement of the lemma.
Obviously, the logarithmic term in the definition of
$\mathcal{S}[\phi]$ does not contribute into the variation.
\end{proof}

Straightforward calculation gives
\[
\varphi_z=\frac{-1}{f'}\left(\frac{f''}{f'}+\frac{1}{\zeta}\right)\circ
f^{-1}(z,t.)
\]
Hence, the action $\mathcal{S}$ can be expressed in terms of the
parametric function $f$ as
\begin{equation}
\mathcal{S}=
\mathcal{S}[f]=\lim\limits_{\varepsilon\to
0}\left\{\int\limits_{U_{\varepsilon}}\bigg|\frac{f''}{f'}+\frac{1}{\zeta}\bigg|^2d\sigma_\zeta+2\pi\log\varepsilon\right\}
+2\pi\log|f'(0,t)|, \label{action3}
\end{equation}
or adding the logarithmic term into the integral we obtain
\begin{equation}
\mathcal{S}[f]=\int\limits_{U}\left(\bigg|\frac{f''}{f'}+\frac{1}{\zeta}\bigg|^2-\frac{1}{|\zeta|^2}\right)d\sigma_\zeta
+2\pi\log|f'(0,t)|. \label{action4}
\end{equation}
Within the quantum theory of Riemann surfaces the Liouville action is a K\"ahler potential of the Weil-Petersson metric on the space of deformations (Teichm\"uller space), see \cite{ZT1, ZT2}.
We use a flat metric instead, nevertheless,
as we show further on, there are several common features between smooth subordination evolution and the Liouville theory. In particular, we shall derive the variation of the logarithmic action $\mathcal S$
and  give connections with the Virasoro algebra and the K\"ahler geometry on the infinite dimensional
manifold $\Diff S^1/S^1$.  

\section{Variation of the logarithmic action and the K\"ahler geometry on $\Diff S^1/S^1$}

We denote the Lie group of $C^{\infty}$ sense preserving
diffeomorphisms of the unit circle $S^1=\partial U$  by $\Diff S^1$.
Each element of $\Diff S^1$ is represented as $z=e^{i\phi(\theta)}$
with a monotone increasing, $C^{\infty}$ real-valued function
$\phi(\theta)$, such that $\phi(\theta+2\pi)=\phi(\theta)+2\pi$.
 The Lie algebra for $\Diff S^1$ is
identified with the Lie algebra $\Vect S^1$ of smooth ($C^{\infty}$)
tangent vector fields to $S^1$ with the Poisson - Lie bracket given
by
$$[\nu_1,\nu_2]={\nu}_1{\nu}'_2-{\nu}_2{\nu}'_1. $$
There is no general theory of infinite dimensional Lie groups,
example of which is under consideration.  The interest to this
particular case comes first of all from the string  theory where the
Virasoro (vertex) algebra appears as the central extension of $\Vect
S^1$ and gives the mode expansion for the energy-momentum tensor. 
The central extension of $\Diff S^1$ is called the Virasoro-Bott group.
Entire necessary background for the construction of the theory
of unitary representations of $\Diff S^1$ is found in the study of
Kirillov's homogeneous K\"ahlerian manifold $\mathcal{M}=\Diff S^1/S^1$. The group
$\Diff S^1$ acts as a group of translations on the manifold $\mathcal{M}$ with
the group $S^1$ as a stabilizer. The K\"ahlerian geometry of $\mathcal{M}$ has been
described by Kirillov and Yuriev in \cite{KY1}. The manifold $\mathcal{M}$
admits several representations, in particular, in the space of
smooth probability measures, symplectic realization in the space of
quadratic differentials. Let  $A$ stand for  the class of all
analytic regular univalent functions $f$ in $U$ normalized by
$f(0)=0$, $f'(0)=1$. We shall use its analytic representation of $\mathcal{M}$
 based on the class $\tilde{A}$ of functions from $A$ which
being extended onto the closure $\overline{U}$ of $U$  are supposed
to be smooth on $S^1$. The class $\tilde{A}$ is dense in $A$ in the
local uniform topology of $U$. There exists a canonical identification
of $\tilde{A}$ with $\mathcal{M}$. As a consequence, $\tilde{A}$ is a homogeneous
space under the left action of $\Diff S^1$, see \cite[Theorem 1.4.1]{AiraultMalliavin} and \cite{Kir}--\cite{KY2}. As it has been mentioned in Section 2, see also \cite{Vasiliev2},
a smooth subodination evolution is governed by the L\"owner-Kufarev equation (\ref{LK})
with a function $p(\zeta,t)$ which may be Schwarz represented by its boundary values
$\rho(e^{i\theta},t)$, such that $4\pi \rho\in \Vect S^1$.

\begin{theorem} Let $z=f(\zeta,t)$ be the parametric function for
the subordination evolution $\Omega(t)$, $t\in[0,\tau)$, $f(\zeta,t)=e^t\zeta+\dots$. Let
$\mathcal{S}[f]$ stand for the logarithmic action. Then,
\begin{equation*}
\frac{d}{dt}\mathcal{S}[f]=\int\limits_{0}^{2\pi}\left[\R\left(1+\frac{e^{i\theta}f''}{f'}\right)\right]^2
\nu(e^{i\theta},t)\,d\theta+
\int\limits_{0}^{2\pi}\R(e^{2i\theta}S_f)\,\nu(e^{i\theta},t)\,d\theta-2\pi,
\end{equation*}
where $\nu\in \Vect S^1$, $\nu>0$ and $\int_0^{2\pi}\nu(e^{i\theta},t)d\theta=4\pi$.
\end{theorem}
\begin{proof}
We start rewriting the expression for $\mathcal{S}[f]$ as
\[
\mathcal{S}[f]=\iint\limits_U\left(\bigg|\frac{f''(\zeta,t)}{f'(\zeta,t)}\bigg|^2+
2\R\frac{f''(\zeta,t)}{\overline{\zeta}f'(\zeta,t)}\right)\,d\sigma_{\zeta}+2\pi t,
\]
and therefore,
\[
\frac{d}{dt}\mathcal{S}[f]=2\R\iint\limits_U\overline{\left(\frac{f''}{f'}+\frac{1}{\zeta}
\right)}\left(\frac{\dot{f}''}{f'}-\frac{f''\dot{f}'}{(f')^2}
\right)\,d\sigma_{\zeta}+2\pi.
\]
Now applying the L\"owner-Kufarev representation $\dot{f}=\zeta f'p(\zeta,t)$, we get
\[\frac{d}{dt}\mathcal{S}[f]=2\R\iint\limits_{U}\overline{\left(\frac{f''}{f'}+\frac{1}{\zeta}
\right)}
\left((1+\zeta\frac{f''}{f'})p(\zeta,t)+\zeta
p'(\zeta,t)\right)'d\sigma_\zeta+2\pi.
\]
In order to apply Green's theorem we remove the singularity at the origin by splitting the integral into
two terms as
\[\frac{d}{dt}\mathcal{S}[f]=2\R\left(\iint\limits_{U_{\varepsilon}}\dots+\iint\limits_{|\zeta|<\varepsilon}\dots\right)+2\pi,
\]
where the second term
\[
\iint\limits_{|\zeta|<\varepsilon}\dots=\int\limits_0^{\varepsilon}\int\limits_0^{2\pi}
\left(\overline{\frac{f''}{f'}}+\frac{e^{i\theta}}{r}\right)(\mbox{holomorphic function})r\,d\theta dr\to 0
\]
as $\varepsilon\to 0$. Applying Green's theorem for the first term and taking to account that $p(0,t)=1$, we obtain
\[
\iint\limits_{U_{\varepsilon}}\dots=\left(\frac{-1}{2i}\int\limits_{S^1}\dots\,d\bar{\zeta}+\frac{1}{2i}\int\limits_{|\zeta|=\varepsilon}
\dots\,d\bar{\zeta}\right)\to \left(\frac{-1}{2i}\int\limits_{S^1}\dots\,d\bar{\zeta}-\pi\right),
\]
as $\varepsilon\to 0$. Thus, we have
\[\frac{d}{dt}\mathcal{S}[f]=\R\int\limits_{0}^{2\pi}\overline{\left(1+e^{i\theta}\frac{f''}{f'}\right)}
\left((1+e^{i\theta}\frac{f''}{f'})p(e^{i\theta},t)+e^{i\theta}
p'(e^{i\theta},t)\right)\,d\theta,
\]
or
\begin{eqnarray*}
\frac{d}{dt}\mathcal{S}[f]=2\pi \int\limits_{0}^{2\pi}\bigg| 1+e^{i\theta}\frac{f''}{f'}\bigg|^2\rho(e^{i\theta},t)\,d\theta&+&
\int\limits_{0}^{2\pi}\R\left(1+e^{i\theta}\frac{f''}{f'}\right)\R
e^{i\theta}p'(e^{i\theta},t)\,d\theta\\&+&\int\limits_{0}^{2\pi}\I\left(1+e^{i\theta}\frac{f''}{f'}\right)\I
e^{i\theta}p'(e^{i\theta},t)\,d\theta.
\end{eqnarray*}
These equalities are thought of as limiting values making use of the
smoothness of $f$ on the boundary. Let us denote by $J_1$, $J_2$ and $J_3$, the first, the second and the third
term respectively in the latter
expression. We have
\[
J_3=\int\limits_{0}^{2\pi}\I\left(1+e^{i\theta}\frac{f''}{f'}\right)\I\int\limits_{0}^{2\pi}
\frac{2e^{i\theta}e^{i\alpha}}{(e^{i\alpha}-e^{i\theta})^2}\rho(e^{i\alpha},t)\,d\alpha\,
d\theta.
\]
Obviously,
\[
\frac{\partial}{\partial\alpha}\left(\frac{e^{i\alpha}+\zeta}{e^{i\alpha}-\zeta}\right)=\frac{-2\,\zeta
i e^{i\alpha}}{(e^{i\alpha}-\zeta)^2}.
\]
Integrating by parts  and applying the Cauchy-Schwarz formula 
we obtain
\[
J_3=2\pi \,\R \int\limits_{0}^{2\pi}\left(e^{i\theta}\frac{f''}{f'}+e^{2i\theta}\left(\frac{f'''}{f'}-\left(\frac{f''}{f'}\right)^2\right)\right)\rho(e^{i\theta},t)\,d\theta.
\]
Using the representation of the function $p(\zeta,t)$ the integral $J_2$ admits the form
\[
J_2=\int\limits_{0}^{2\pi}\R\left(1+e^{i\theta}\frac{f''}{f'}\right)\R\int\limits_{0}^{2\pi}
\frac{2e^{i\theta}e^{i\alpha}}{(e^{i\alpha}-e^{i\theta})^2}\rho(e^{i\alpha},t)\,d\alpha\,
d\theta.
\]
Changing the order of integration implies
\begin{eqnarray*}
J_2&=&\int\limits_{0}^{2\pi}\R\int\limits_{0}^{2\pi}\R\left(1+e^{i\theta}\frac{f''}{f'}\right)
\frac{2e^{i\theta}e^{i\alpha}}{(e^{i\alpha}-e^{i\theta})^2}\rho(e^{i\alpha},t)\,d\alpha\,
d\theta\\ &=&
\R\int\limits_{0}^{2\pi}\rho(e^{i\alpha},t)\left(\int\limits_{0}^{2\pi}\R\left(1+e^{i\theta}\frac{f''}{f'}\right)
\frac{2e^{i\theta}e^{i\alpha}}{(e^{i\alpha}-e^{i\theta})^2}\,d\theta\right)
d\alpha.
\end{eqnarray*}
Integrating by parts we obtain
\[
J_2=\int\limits_{0}^{2\pi}\rho(e^{i\alpha},t)\left(\R (-i)\int\limits_{0}^{2\pi}\frac{\partial}{\partial \theta}\R\left(1+e^{i\theta}\frac{f''}{f'}\right)
\frac{e^{i\theta}+e^{i\alpha}}{e^{i\alpha}-e^{i\theta}}\,d\theta\right)
d\alpha.
\]
The internal integral represents an analytic function by the Cauchy formula (modulo an imaginary
constant). Taking into account the normalization at the origin we get
\[
J_2=2\pi\, \R \int\limits_{0}^{2\pi}\left(e^{i\alpha}\frac{f''}{f'}+e^{2i\alpha}\left(\frac{f'''}{f'}-\left(\frac{f''}{f'}\right)^2\right)\right)\rho(e^{i\alpha},t)\,d\alpha=J_3.
\]
Summing up $J_1+J_2+J_3$, and taking into account $\nu=4\pi\,\rho$, concludes the proof.  
\end{proof}

In the particular case of the Laplacian growth evolution this theorem has been proved in \cite{Vasiliev3}.
The normal velocity of the boundary is equal to the gradient of the Green function and $\nu(e^{i\theta},t)=
2/|f'(e^{i\theta},t)|^2$.

In two-dimensional conformal field theories \cite{Goddard}, the algebra of energy momentum tensor is deformed by a central extension due to the conformal anomaly and becomes the {\it Virasoro algebra}.
The Virasoro algebra is spanned
by elements $e_k=\zeta^{1+k}\partial$, $k\in \mathbb Z$ and $c$ with $e_{k}+e_{-k}$, where $c$ is a real number, called the central charge, and the Lie brackets are defined by
\[
[e_m,e_n]_{Vir}=(n-m)e_{m+n}+\frac{c}{12}m(m^2-1)\delta_{n,-m},\quad [c, L_k]=0.
\]
The Virasoro algebra ($Vir$) can be realized as a central extension
of $\Vect S^1$ by defining 
\[
[\phi\partial+ca , \psi\partial+cb]_{Vir}= (\phi\psi'-\phi'\psi)\partial+\frac{c}{12}\omega(\phi,\psi),
\]
(whereas $[\phi, \psi]= \phi\psi'-\phi'\psi$), where the bilinear antisymmetric form $\omega(\phi,\psi)$ on $\Vect S^1$ is given by
\[
\omega(\phi,\psi)=-\frac{1}{4\pi}\int\limits_{0}^{2\pi}(\phi'+\phi''')\psi d\theta,
\]
and $a,b$ are numbers. This form defines the  Gelfand-Fuks cocycle on $\Vect S^1$ and satisfies the Jacobi identity. The factor of 1/12 is merely a matter of convention.
The manifold $\mathcal{M}$ being considered as a realization $\tilde{A}$ admits affine coordinates
$\{c_2,c_3,\dots\}$, where $c_k$ is the $k$-th coefficient of a univalent functions $f\in \tilde{A}$.
Due to de Branges' theorem \cite{Branges}, $\mathcal{M}$ is a bounded open subset of $\{|c_k|<k+\varepsilon\}$.

The Goluzin-Schiffer variational formula lifts the actions from the
Lie algebra $\Vect S^1$ onto $\tilde{A}$. Let $f\in\tilde{A}$ and
let $\nu(e^{i\theta})$ be a $C^{\infty}$ real-valued function in
$\theta\in(0,2\pi]$ from $\Vect S^1$ making an infinitesimal action
as $\theta \mapsto \theta+\tau \nu(e^{i\theta})$. Let us consider a
variation of $f$ given by
\begin{equation}
L_{\nu}[f](\zeta)=-\frac{f^2(\zeta)}{2\pi
i}\int\limits_{S^1}\left(\frac{wf'(w)}{f(w)}\right)^2\frac{\nu
(w)}{f(w)-f(\zeta)}\frac{dw}{w} .\label{var}
\end{equation}
 Kirillov and Yuriev \cite{KY1,KY2} have established
that the variations $L_{\nu}[f](\zeta)$ are closed with respect
to the commutator and the induced Lie algebra is the same as $\Vect
S^1$. Moreover, Kirillov's result \cite{Kir} states that there is
the exponential map $\Vect S^1\to \Diff S^1$ such that the subgroup
$S^1$ coincides with the stabilizer of the map $f(\zeta)\equiv
\zeta$ from $\tilde{A}$.

It is convenient \cite{Kir98} to extend (\ref{var}) by complex linearity to $\mathbb C\Vect S^1\to \Vect \tilde{A}$.
Taking $\nu_k=-ie^{ik\theta}$, $k\geq 0$ from the basis of $\mathbb C\Vect S^1$, we obtain the expressions for $L_k=\delta_{\nu} f$, 
$f\in \tilde{A}\simeq \mathcal{M}$ (see formula (\ref{var})),  as
\[
L_0=\zeta f'(\zeta)-f(\zeta), \quad L_k=\zeta^{1+k}f'.
\]
The computation of $L_k$ for $k<0$ is more difficult because poles of the integrant.
For example,
\[
L_{-1}=f'-1-2c_2f,\quad L_{-2}=\frac{f'}{\zeta}-\frac{1}{f}-3c_2+(c_2^2-4c_3)f,
\]
(see, e.g., \cite{Kir98}). In terms of the coordinates $\{c_2,c_3,\dots\}$ on $\mathcal{M}$ 
\[
L_k=\partial_k+\sum\limits_{n=1}^{\infty}(n+1)c_n\partial_{k+n},\quad L_0=\sum\limits_{n=1}^{\infty}nc_n\partial_{n},
\]
for $k>0$, where $\partial_{k}=\partial/\partial c_{k+1}$.

Neretin \cite{Neretin} introduced the sequence of polynomials $P_k$, in the coordinates $\{c_2,c_3,\dots\}$ on $\mathcal{M}$ by the following recurrent relations
\[
L_m(P_n)=(n+m)P_{n-m}+\frac{c}{12}m(m^2-1)\delta_{n,m},\quad P_0=P_1\equiv 0,\quad P_k(0)=0,
\]
where the central charge $c$ is fixed. This gives, for example, $P_2=\frac{c}{2}(c_3-c_2^2)$,
$P_3=2c(c_4-2c_2c_3+c_2^3)$. In general, the polynomials $P_k$ are homogeneous with respect to
rotations of the function $f$. It is worthy to mention that estimates of the absolute value of these polynomials
has been a subject of investigations in the theory of univalent functions for a long time, e.g., for $|P_2|$ we have
$|c_3-c_2^2|\leq 1$ (Bieberbach 1916 \cite{Bieberbach}), for estimates of $|P_3|$ see \cite{Gromova, Lehto, Tammi1, Tammi2}. For the Neretin polynomials one can construct the generatrix function  
\[
P(\zeta)=\sum\limits_{k=1}^{\infty}P_k\zeta^k=\frac{c\zeta^2}{12}S_f(\zeta),
\]
where $S_f(\zeta)$ is the Schwarzian derivative of $f$,
Let $\nu\in \mathbb{C}\Vect S^1$ and $\nu^g$ be the associated right-invariant tangent vector field defined at $g\in \Diff S^1$. For the basis $\nu_k=-ie^{ik\theta}\partial$, one constructs the corresponding  associated right-invariant basis $\nu_k^g$. By $\{\psi_{-k}\}$ we denote the dual basis of 1-forms such that the value
of each form on the vector $\nu_k^g$ is given as
\[
(\psi_k, \nu_n^g)=\delta_{k+n,0}.
\]
Let us construct the 1-form $\Psi$ on $\Diff S^1$ by
\[
\Psi=\sum\limits_{k=1}^{\infty}(P_k\circ \pi)\psi_k,
\]
where $\pi$ means the natural projection $\Diff S^1\to \mathcal{M}$.
This form appeared in \cite{AiraultMalliavin, Airault2}  in the context of the construction of a unitarizing probability measure for the Neretin representation of $\mathcal{M}$. It is invariant under the left action of $S^1$. If $f\in \tilde{A}$ represents $g$ and $\nu\in \Vect S^1$, then the value of the form $\Psi$ on the vector $\nu$ is
\[
(\Psi, \nu)_{f}=\int\limits_{0}^{2\pi}e^{2i\theta}\nu(e^{i\theta}) S_f \,\,
d\theta,
\]
see \cite{AiraultMalliavin, Airault2}. So the variation of the logarithmic action given in Theorem 1 becomes
\begin{equation*}
\frac{d}{dt}\mathcal{S}[f]=\int\limits_{0}^{2\pi}\left[\R\left(1+\frac{e^{i\theta}f''}{f'}\right)\right]^2
\nu(e^{i\theta},t)\,d\theta+
\R (\Psi, \nu)_{f}-2\pi.
\end{equation*}
Taking into account the definition of the mean curvature $\varkappa(z,t)$ of the boundary of $\Omega(t)$, and the
normal velocity $v_n$, we conclude that
\begin{equation*}
\frac{d}{dt}\mathcal{S}[f]=4\pi\int\limits_{\partial \Omega(t)} (\varkappa v_n)^2|dz|+
\R (\Psi, \nu)_{f}-2\pi.
\end{equation*}

\end{document}